\documentclass[superscriptaddress,amsmath, nofootinbib]{revtex4}
\usepackage{amsfonts}
\usepackage{graphicx}
\usepackage[brazil]{babel}
\usepackage[latin1]{inputenc}
\usepackage{amsmath}
\usepackage{amssymb}
\begin{document}


\title{Geodesic motion on the symplectic leaf of $SO(3)$ with distorted $e(3)$ algebra and Liouville integrability of a free rigid body.}

\author{Alexei A. Deriglazov }
\email{alexei.deriglazov@ufjf.br} \affiliation{Depto. de Matem\'atica, ICE, Universidade Federal de Juiz de Fora,
MG, Brazil} 

\date{\today}

\begin{abstract}
The solutions to the Euler-Poisson equations are geodesic lines of $SO(3)$ manifold with the metric determined by inertia tensor. However, the Poisson structure on the corresponding symplectic leaf does not depend on the inertia tensor. We calculate its explicit form and confirm that it differs from the algebra $e(3)$. The obtained Poisson brackets are used to demonstrate the Liouville integrability of a free rigid body. The general solution to the Euler-Poisson equations is written in terms of exponential of the Hamiltonian vector field.  
\end{abstract}   

\maketitle 



\section{Introduction: Lagrangian of a rigid body in terms of unconstrained rotation vector.}

In the previous work \cite{AAD23} we discussed the dynamics of a rigid body, taking its Lagrangian action 
\begin{eqnarray}\label{u1.1}
S=\int dt ~\frac12\sum_{N=1}^{n}m_N\dot{\bf y}_N^2+\frac12\sum_{A=2}^{4}\sum_{N=2}^{n}\lambda_{AN}\left[({\bf y}_A-{\bf y}_1, {\bf y}_N-{\bf y}_1)-a_{AN}\right], 
\end{eqnarray}
as the only starting point of the analysis.
Here ${\bf y}_N=(y^1_N, y^2_N, y^3_N)$, $N=1, 2, \ldots , n$ are coordinates of $n$ particles of the body, and $\lambda_{AN}$ are the Lagrangian multipliers that take into account the constraints among the body's particles (we follow the notation of the work \cite{AAD23}). 
Assuming the expression (\ref{u1.1}), we no longer need any additional postulates or assumptions about the behavior of the rigid body. All the basic quantities and characteristics of a rigid body, as well as the equations of motion and integrals of motion, are obtained from the variational problem by direct and unequivocal calculations within the framework of standard methods of classical mechanics.  This can be compared with the standard approach \cite{Mac_1936,Lei_1965,Landau_8}, where a number of postulates should be assumed: on the behavior of the center of mass, as well as on  the conservation of energy and angular momentum. Herewith some important properties of the theory not always are taken into account, see \cite{AAD23}  for the details. 

The analysis of equations following from the action (\ref{u1.1}) shows that all their solutions are of the form
\begin{eqnarray}\label{u1.2}
{\bf y}_N(t)={\bf C}_0+{\bf V}_0 t+{\bf x}_N(t), \qquad \mbox{where} \quad   {\bf x}_N(t)=R(t){\bf x}_N(0). 
\end{eqnarray}
Here the term
${\bf C}_0+{\bf V}_0 t$ describe the motion of the center of mass, while the last term describe the motion of body's particle with the coordinates $x^i_N(t)$ determined with respect to the center of mass. $R_{ij}(t)$ is an orthogonal matrix, $R^TR={\bf 1}$, $\det R=+1$, that determines this rotational movement. The dynamics of $R_{ij}$ is completely determined by its own Lagrangian action
\begin{eqnarray}\label{u1.3}
S=\int dt ~ ~ \frac12 g_{ij}\dot R_{ki}\dot R_{kj} -\frac12 \lambda_{ij}\left[R_{ki}R_{kj}-\delta_{ij}\right],
\end{eqnarray}
with the universal initial conditions $R_{ij}(0)=\delta_{ij}$, implied by Eq. (\ref{u1.2}). Hamiltonian formulation of the theory (\ref{u1.3}) can be constructed using the phase space with mutually independent variables $R_{ij}(t)$ and $\Omega_i(t)$, the latters represent the Hamiltonian counterpart of angular velocity in the body. Their Hamiltonian equations turn out to be just the Euler-Poisson equations \cite{AAD23}
\begin{eqnarray}\label{u1.4} 
\dot R_{ij}=-\epsilon_{jkm}\Omega_k R_{im}, 
\end{eqnarray}
\begin{eqnarray}\label{u1.5}
I\dot{\boldsymbol\Omega}=[I{\boldsymbol\Omega}, {\boldsymbol\Omega}],   
\end{eqnarray}
where $I_{ij}$ are components of the inertia tensor.  These equations of motion are still written for an excess number of variables. For any solution to Eqs. (\ref{u1.4}) and (\ref{u1.5}) with above mentioned initial conditions, the nine matrix elements $R_{ij}(t)$ obey to six 
constraints $R^T(t)R(t)=1$, so we need to know only some $9-6=3$ independent parameters to specify the matrix $R$. 

\begin{figure}[t] \centering
\includegraphics[width=06cm]{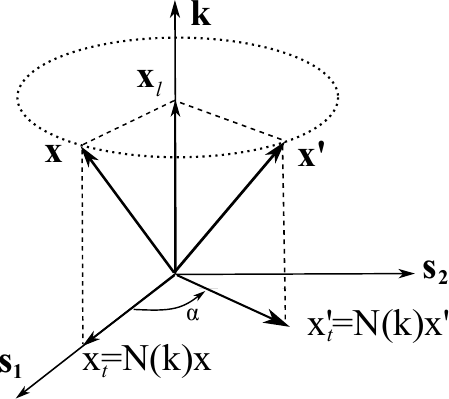}
\caption{The rotation matrix $R_{ij}$ can be parameterized by the unit vector ${\bf k}$ and the angle $\alpha<\pi$.}\label{RB_1}
\end{figure}

There are many different ways to parameterize the rotation matrices \cite{Dub_2001,Arn_2,Fom_2004,Arn_1}. 
In this work we use the parameterization defined with help of the rotation, that can be unambiguously associated with each element $R_{ij}$ of $SO(3)$ as follows. The 
equation $\det(R-\lambda{\bf 1})=0$ for determining eigenvalues of $R$ always admits $\lambda=1$ as a solution. Indeed, with $\lambda=1$ we 
have: $\det(R-{\bf 1})=\det(R-RR^T)=\det R\det({\bf 1}-R^T)=-\det(R-{\bf 1})$, which implies  $\det(R-{\bf 1})=0$. Then the equation $R{\bf k}={\bf k}$ for eigenvectors has two solutions, say ${\bf k}_1$ and ${\bf k}_2$, where ${\bf k}_1$ and ${\bf k}_2$ are unit vectors in opposite directions. They determine the axis of rotation, the points of which remain fixed under the transformation generated by $R$ in ${\mathbb R}^3$. So the transformation $R: x^i\rightarrow x'^i=R_{ij}x^j$ can be considered as a rotation of the spatial points around this axis through some angle $0\le\alpha<\pi$, see Figure \ref{RB_1}.  
Between the vectors ${\bf k}_1$ and ${\bf k}_2$, we choose the one for which this rotation occurs counterclockwise when viewed from the end of this vector. 
Let us call this unit vector ${\bf k}$. Expressing $x'^i$ in terms of $x^j$, ${\bf k}$ and $\alpha$, we get the matrix $R$ in terms of ${\bf k}$ and $\alpha$. To this aim,  we use the projectors: $\delta_{ij}=P_{ij}({\bf k})+N_{ij}({\bf k})$, $P_{ij}({\bf k})\equiv k_i k_j$, $N_{ij}({\bf k})\equiv\delta_{ij}-k_i k_j$ to decompose the vectors $x^i$ and $x'^i$ on transverse and longitudinal parts with respect to the vector ${\bf k}$ 
\begin{eqnarray}\label{u1.6}
x^i=x^i_l+x^i_t=P_{ij}x^j+N_{ij}x^j, \qquad 
x'^i=x'^i_l+x'^i_t=P_{ij}x'^j+N_{ij}x'^j.
\end{eqnarray}
On the plane othogonal to ${\bf k}$ we define the basis composed of the vectors ${\bf s}_1={\bf x}_t/|{\bf x}_t|$, 
${\bf s}_2=[{\bf k}, {\bf s}_1]=[{\bf k}, {\bf x}_t]/|{\bf x}_t|$.  Then  
${\bf x'}_t={\bf s}_1|{\bf x'}_t|\cos\alpha+{\bf s}_2|{\bf x'}_t|\sin\alpha=N{\bf x}\cos\alpha+[{\bf k}, {\bf x}]\sin\alpha$, so ${\bf x}'=P{\bf x}+N{\bf x}\cos\alpha+[{\bf k}, {\bf x}]\sin\alpha$. This gives us the matrix $R$ 
\begin{eqnarray}\label{u1.7}
R_{ij}=\delta_{ij}\cos\alpha+(1-\cos\alpha)k_i k_j-\epsilon_{ijk}k_k\sin\alpha. 
\end{eqnarray}
Further, using the identities $\sin\alpha=2\tan^2(\alpha/2)/(1+\tan^2(\alpha/2)$ and $\cos\alpha=(1-\tan^2(\alpha/2))/(1+\tan^2(\alpha/2)$, we introduce the vector 
\begin{eqnarray}\label{u1.8}
n_i=k_i\tan(\alpha/2), \qquad 0\le{\bf n}^2=\tan^2(\alpha/2)<\infty.  
\end{eqnarray}
In terms of this unconstrained vector, we get the final form of the desired parameterization\footnote{The rotation vector is related with Cayley-Klein (or quaternion) parameters $q^0, {\bf q}$ as follows: $n^i=q^i/q^0$.} 
\begin{eqnarray}\label{u1.9}
R_{ij}({\bf n})=\frac{1}{1+{\bf n}^2}\left[(1-{\bf n}^2)\delta_{ij}+2n_i n_j-2\epsilon_{ijk}n_k\right]. 
\end{eqnarray}

The unit element of $SO(3)$ corresponds to the values $n^i=0$ of the parameters: $R_{ij}(0)=\delta_{ij}$. We emphasis that working with the Euler-Poisson equations, we are interested in trajectories that pass through the unit of $SO(3)$. In this respect, the coordinates $n_i$ are more convenient than the Euler angles, since in the latter case the unit lies outside the Euler coordinate system which can lead to misunderstanding, 
see \cite{AAD23_2}. 

To express the Lagrangian (\ref{u1.3}) in terms of ${\bf n}$, it is convenient first to rewrite it in terms of angular velocity\footnote{The variational problems (\ref{u1.3}) and (\ref{u1.10}) are equivalent, see the Appendix in \cite{AAD23}.} in the body as follows:
\begin{eqnarray}\label{u1.10}
L=\frac12 I_{ij}\Omega_i\Omega_j -\frac12 \lambda_{ij}\left[R_{ki}R_{kj}-\delta_{ij}\right], \qquad \mbox{where} \quad \Omega_k\equiv-\frac12\epsilon_{kij}(R^T\dot R)_{ij}. 
\end{eqnarray}
Using the parameterization (\ref{u1.9}) we get
\begin{eqnarray}\label{u1.11}
\Omega_k({\bf n}, \dot{\bf n})=2(A^T\dot{\bf n})_k, 
\end{eqnarray}
where the conversion matrix is 
\begin{eqnarray}\label{u1.12}
A_{ij}=\frac{1}{1+{\bf n}^2}\left[\delta_{ij}-\epsilon_{ijk}n_k\right], \quad \mbox{then} \quad A^{-1}_{ij}\equiv\tilde A_{ij}=\delta_{ij}+n_j n_j
+\epsilon_{ijk}n_k. 
\end{eqnarray}
The rotation matrix $R$ can be written in terms of these matrices as follows: $R=A\tilde A^T$. 
Using these expressions in (\ref{u1.10}), we get the Lagrangian in terms of unconstrained variables
\begin{eqnarray}\label{u1.13}
L=2 I_{ij}(A^T\dot{\bf n})_i(A^T\dot{\bf n})_j=\frac12G_{ij}({\bf n})\dot n_i\dot n_j=\frac{2}{(1+{\bf n}^2)^2}I_{ij} \left[\dot{\bf n}-[{\bf n}, \dot{\bf n}]\right]_i
\left[\dot{\bf n}-[{\bf n}, \dot{\bf n}]\right]_j. 
\end{eqnarray}
This can be considered as describing a geodesic motion\footnote{More exactly, this Lagrangian implies geodesic equations in the natural parametrization, see Sect. 6.5 in \cite{deriglazov2010classical}.} of the particle ${\bf n}$ in three-dimensional space with the metric $G_{ij}({\bf n})=4(AIA^T)_{ij}$.  The geodesics that pass through the origin describe the possible motions of the rigid body. The metric has an unusual asymptotic behavior: $G\rightarrow\delta$ as ${\bf n}\rightarrow 0$, and $G\rightarrow 1/{\bf n}^2$ as ${\bf n}\rightarrow\infty$,  that is this is almost Euclidean in the vicinity of origin while vanishes at infinity. Note that ${\bf n}\rightarrow\infty$ corresponds to the angle $\alpha\rightarrow\pi$. As our aim here is to study the algebraic properties of Hamiltonian quantities of the theory, this coordinate singularity does not represent any special problem.

\section{Canonical Hamiltonian formulation and integrability.} 
The Hamiltonian formulation immediately follows from the expression (\ref{u1.13}). Computing the conjugate momenta $\pi_i=\partial L/\partial\dot n^i$, we get
\begin{eqnarray}\label{u1.14}
\pi_i=G_{ij}\dot n_j, \qquad \mbox{then} \quad \dot n_i=G^{-1}_{ij}\pi_j, \quad \mbox{where} \quad G^{-1}=\frac14 \tilde A^TI^{-1}\tilde A. 
\end{eqnarray}
Then the Hamiltonian $H=\pi_i\dot n_i-L$ is
\begin{eqnarray}\label{u1.15}
H=\frac12 G^{-1}_{ij}\pi_i \pi_j=\frac18 I^{-1}_{ij}(\tilde A{\boldsymbol\pi})_i(\tilde A{\boldsymbol\pi})_j=\frac18 I^{-1}_{ij}[{\boldsymbol\pi}+({\bf n}, {\boldsymbol\pi}){\bf n}+[{\boldsymbol\pi}, {\bf n}]]_i[{\boldsymbol\pi}+({\bf n}, {\boldsymbol\pi}){\bf n}+[{\boldsymbol\pi}, {\bf n}]]_j,
\end{eqnarray}
while the canonical Poisson brackets are 
\begin{eqnarray}\label{u1.15.1}
\{n_i, \pi_j\}=\delta_{ij}, \qquad \{n_i, n_j\}=\{\pi_i, \pi_j\}=0.   
\end{eqnarray}
For the latter use we observe, that they imply
\begin{eqnarray}\label{u1.18}
\{(\tilde A^T{\boldsymbol\pi})_i, (\tilde A{\boldsymbol\pi})_j\}=0. 
\end{eqnarray}
The Hamiltonian equations of motion can be then obtained according the standard rule: $\dot n_i=\{ n_i, H\}$, $\dot\pi_i=\{\pi_i, H\}$, and read as follows: 
\begin{eqnarray}\label{u1.16}
\dot n_i=\frac14(\tilde A^T I^{-1}\tilde A{\boldsymbol\pi})_i, \qquad 
\dot\pi_i=-\frac14[\delta_{ij}({\bf n}, {\boldsymbol\pi})+\pi_in_j+\epsilon_{ijk}n_k](I^{-1}\tilde A{\boldsymbol\pi})_j.
\end{eqnarray}
Equivalently, they can be obtained as the conditions of extremum of the first-order Hamiltonian action
\begin{eqnarray}\label{u1.16.1}
S_H=\int dt  ~\pi_i\dot n_i-H(n_i, \pi_j).
\end{eqnarray}

Let us discuss the integrability of the rigid body equations (\ref{u1.16}). They admite four classical integrals of motion.
The first is the energy $H$ given in Eq. (\ref{u1.15}). Three more integals of motion\footnote{On the subset of solutions which describe the moviments of a body (they are the solutions that pass through unit element of $SO(3)$), the four integrals are not independent: $E=I^{-1}_{ij} m_i m_j$, see \cite{AAD23} for the details.} are the components of angular momentum of the body: ${\bf m}=RI{\boldsymbol\Omega}$. To see this in the Hamiltonian framework, we write them in terms of canonical variables
\begin{eqnarray}\label{u1.17}
m_i=\frac12(\tilde A^T{\boldsymbol\pi})_i=\frac12(\pi_i+({\bf n}{\boldsymbol\pi})n_i  -[{\boldsymbol\pi}, {\bf n}]_i). 
\end{eqnarray}
Remarkably, the functions $m_i(n_j, \pi_k)$ (and hence their brackets)  do not depend on the inertia tensor. 
Using  Eqs. (\ref{u1.15}), (\ref{u1.18}) and (\ref{u1.17}), we immediately conclude 
\begin{eqnarray}\label{u1.19}
\{ m_i, H \}=0,
\end{eqnarray}
which implies the conservation of the angular momentum: $\dot m_i=\{ m_i, H \}=0$. 

The equalities (\ref{u1.17})  are invertible with respect to $\pi_i$, so we can work with the rigid body in terms of non canonical phase-space variables $n_i$, $m_j$ instead of $n_i$, $\pi_j$. Making this change of variables in (\ref{u1.16}), we get an equivalent system\footnote{The Hamiltonian in these variables 
is $\frac12(RI^{-1}R^T)_{ij}m_im_j$, see also Eq. (\ref{u1.25}) below.}  
\begin{eqnarray}\label{u1.20}
\dot n_i=\frac12(\tilde A^T I^{-1}R^T{\bf m})_i, \qquad  \dot m_i=0. 
\end{eqnarray}
By the way, we reduced the number of differential equations from six to three, the latters contain now three integration constants $m_i$. Poisson structure of the theory in these variables is 
\begin{eqnarray}\label{u1.21}
\{n_i, n_j\}=0, \qquad   \{n_i, m_j\}=\frac12 \tilde A_{ij}=\frac12 [\epsilon_{ijk}n_k+\delta_{ij}+n_i n_j], \cr 
\{m_i, m_j\}=\frac{1}{1+{\bf n}^2}[\epsilon_{ijk}m_k+
(\hat n_{ij}n_k+cycle\,(ijk) )m_k], \quad 
\end{eqnarray}
where $\hat n_{ij}\equiv \epsilon_{ijp}n_p$ is the antisymmetric matrix equivalent to the vector $n_p$. The second equality shows that the conversion matrix $\tilde A$ represents the off-diagonal part of the Poisson tensor in these coordinates. 
Even in the linear approximation, this algebra is different from the isometry algebra $e(3)$ of ${\mathbb R}^3$. We also emphasise that the Poisson tensor determined by (\ref{u1.21}) is just the canonical Poisson tensor (\ref{u1.18}) written in a noncanonical coordinates of the phase space. This implies, in particular, that the Poisson structure (\ref{u1.21}) is nondegenerate and hence does not admite the Casimir functions. 

The expression $(\hat n_{ij}n_k+cycle\,(ijk) )$ is antisymmetric on $k$ and $j$, so we have the identity
\begin{eqnarray}\label{u1.22}
(\hat n_{ij}n_k+cycle\,(ijk) )m_k m_j=0.  
\end{eqnarray}
Using this identity and Eq. (\ref{u1.21}) we conclude, that the phase-space function ${\bf m}^2$ has vanishing brackets with $m_i$
\begin{eqnarray}\label{u1.23}
\{m_i, {\bf m}^2\}=0.
\end{eqnarray}
This equation together with (\ref{u1.19}) show that the three independent integrals of motion $H$, ${\bf m}^2$ and $m_3$ are in involution. According to the Liouville's theorem \cite{Arn_1, Dub_2001,Fom_2004}, this implies the integrability in quadratures of equations of motion of the free rigid body.

An advantage of the unconstrained problem (\ref{u1.15}) is that we can use now the known formula of Hamiltonian mechanics to write the solution to the equations (\ref{u1.16}) in terms of exponential of the Hamiltonian vector field \cite{AAD_2022}
\begin{eqnarray}\label{u1.23.1}
z^a(t, z^b_0)=e^{t\{z^c_0, ~H(n_{0 i}, \pi_{0 j})\}\frac{\partial}{\partial z^c_0}}z^a_0, \qquad \mbox{where} \quad   z^a=( n_i, \pi_j). 
\end{eqnarray}
After computing all derivatives, one should take $n_{0i}=0$, in accordance with the initial conditions.The resulting expression depends on $3$ arbitrary constants $\pi_i$, and is therefore a general solution to the Euler-Poisson equations.

\section{Hamiltonian formulation in terms of phase-space variables $n_i$, $\Omega_j$.}  To describe the speed of rotation in the theory of a rigid body, several interrelated variables are used: angular velocity $\omega_i$, angular velocity in the body $\Omega_i$, angular momentum $m_i$, and angular momentum in the body $M_i$. The relations between them are 
\begin{eqnarray}\label{u1.24}
2(A^T\dot{\bf n})_i={\bf\Omega}=R^T{\boldsymbol\omega}=I^{-1}R^T{\bf m}=I^{-1}{\bf M}, 
\end{eqnarray}
see \cite{AAD23} for the details. According to this, the kinetic part of the Lagrangian (\ref{u1.10}) can be presented in various forms as follows:
\begin{eqnarray}\label{u1.25}
E=\frac12I_{ij}\Omega_i\Omega_j=\frac12(RI^{-1}R^T)_{ij}m_im_j=\frac12 (RIR^T)_{ij}\omega_i\omega_j=\frac12I^{-1}_{ij}M_i M_j.
\end{eqnarray}
All the basic quantities in (\ref{u1.24}) are related by invertible matrices, so any one of them can be used instead of the canonical momentum ${\boldsymbol\pi}$
in the Hamiltonian formalism. The most simple form of the Lagrangian is achieved in terms of $\Omega_i$ or $M_i$. So it is interesting to consider the Hamiltonian formulation in terms of one of these variables. Let us consider the case of $\Omega_i$.  Using Eqs. (\ref{u1.11}) and (\ref{u1.14}) we get 
\begin{eqnarray}\label{u1.26}
{\boldsymbol\Omega}=\frac12 I^{-1}\tilde A{\boldsymbol\pi}, \qquad {\boldsymbol\pi}=2AI{\boldsymbol\Omega}. 
\end{eqnarray}
Using this in Eq. (\ref{u1.15}) we get the Hamiltonian 
\begin{eqnarray}\label{u1.27}
H=\frac12 I_{ij}\Omega_i\Omega_j,
\end{eqnarray}
while the canonical Poisson brackets (\ref{u1.15.1}) imply
\begin{eqnarray}\label{u1.28}
\{n_i, n_j\}=0, \qquad \{n_i, \Omega_j\}=\frac12(\tilde A^TI^{-1})_{ij}=\frac12[-\epsilon_{ikp}n_p+\delta_{ik}+n_i n_k]I^{-1}_{kj},  \quad ~ \cr 
\{\Omega_i, \Omega_j\}=\frac{-1}{1+{\bf n}^2}(I^{-1})_{ia}(I^{-1})_{jb}\left[\epsilon_{abc}(I\Omega)_c+(\hat n_{ab}n_c+cycle \, (abc)\,)(I\Omega)_c\right]. 
\end{eqnarray}
If we take $M_i$ instead of $\Omega_i$, we obtain the similar expressions, but without  the inertia tensor $I$. Any case, the Poisson structure is nondegenerate and even in the linear approximation differs from the algebra $e(3)$. 

There is the identity
\begin{eqnarray}\label{u1.29}
(\hat n_{ab}n_c+cycle \, (abc)\,)(I\Omega)_c \Omega_b=-{\bf n}^2[I{\boldsymbol\Omega}, {\boldsymbol\Omega}]_a.
\end{eqnarray}
To prove this, we observe that it is $SO(3)$ covariant equation, so we can assume that the inertia tensor is of diagonal form, $I_{ij}=diagonal\, (I_1, I_2, I_3)$. Then the identity can be easily confirmed by direct calculation. 

Using Eqs. (\ref{u1.27}), (\ref{u1.28}) and (\ref{u1.29}), we obtain Hamiltonian equations of motion of a rigid body in terms of these variables
\begin{eqnarray}\label{u1.30}
\dot n_i=\frac12(\tilde A^T\Omega)_i, \qquad I_{ij}\dot\Omega_j=[I{\boldsymbol\Omega}, ~ {\boldsymbol\Omega}]_i. 
\end{eqnarray}
As it should be expected, the equations for $\Omega_i$  are just the Euler equations.

\section{Conclusion.}
According to classical mechanics \cite{Arn_1,deriglazov2010classical}, any mechanical system with kinematic constraints, when rewritten through the  unconstrained variables, looks like the geodesically moving particle in a curved space.  In this work we have done this for the case of an asymmetric rigid body, obtaining the explicit form of the resulting metric (\ref{u1.13}) in terms of unconstrained variables (\ref{u1.8}). It should be noted that in the case of Euler angles, the metric has more or less simple form only for the symmetric top \cite{Arn_1}. Further, any mechanical system with unconstrained configuration variables leads to a symplectic manifold equipped with the canonical Poisson bracket. For the rigid body, we have done this first in terms of canonical variables $n_i, \pi_j$ with the canoinical Poisson brackets (\ref{u1.15.1}), then in terms of $n_i$ and angular momentum $m_j$ with the Poisson brackets (\ref{u1.21}), and at last in terms of $n_i$ and angular velocity $\Omega_j$ with the Poisson brackets (\ref{u1.28}). Using the obtained brackets, it is easy to confirm that the theory admits three integrals of motion in involution, and therefore is integrable according to Liouville. The general solution to the Euler-Poisson equations (\ref{u1.4}), (\ref{u1.5}) is written in terms of exponential of the Hamiltonian vector field in (\ref{u1.23.1}). 

According to the Dirac's canonical quantization paradigm, the Poisson structure is a determining factor in the construction of semiclassical models of the relativistic spin and massless polarized particles \cite{AAD_2019,AAD_2021_1}, and should resemble the algebra of quantum observables. For the massive spinning particle, although a formulation based on reducible variables is possible \cite{Han_1974}, however, for independent variables the construction becomes more simple and transparent \cite{deriglazov2010classical,AAD_2019}. 
Another important point is that the known models of spinning particles are direct analogues of the totally symmetric body. As a result, the evolution of spin of a free particle turns out to be trivial. This could be changed by constructing a model that would be analogous to an asymmetric body, whose rotation around the intermediate axis of inertia is known to be unstable. This could expand the range of applicability of the semiclassical models, since the instability would mimic the quantization of spin in an external magnetic field \cite{AAD_2019}.

In conclusion we note the following. It is sometimes stated in the literature that the Poisson structure associated with the rigid body is closely related with the symmetry algebra $e(3)$ of ${\mathbb R}^3$.  However, the Poisson brackets (\ref{u1.21}) and (\ref{u1.28}), obtained above, are non-degenerate and even in the linear approximation are different from $e(3)$.

\begin{acknowledgments}
The work has been supported by the Brazilian foundation CNPq (Conselho Nacional de
Desenvolvimento Cient\'ifico e Tecnol\'ogico - Brasil). 
\end{acknowledgments}


\begin{thebibliography}{99}

\bibitem{Mac_1936} W. D. MacMillan, {\it Dynamics of rigid bodies}, (Dover Publications Inc., New-York, 1936). 

\bibitem{Lei_1965} E. Leimanis, {\it The general problem of the motion of coupled rigid bodies about a fixed point}, (Springer-Verlag, 1965). 

\bibitem{Landau_8}
L. D. Landau and E. M. Lifshitz, {\it Mechanics}, Volume 1, third edition, (Elsevier, 1976).

\bibitem{Dub_2001} B. A. Dubrovin, I. M. Krichever and S. P. Novikov, {\it Integrable Systems I}, in: V. I. Arnold, {\it Dynamical systems III}, (Springer-Verlag, 2001). 

\bibitem{Arn_2} V. I. Arnold, V. V. Kozlov and A. I. Neishtadt, {\it Mathematical aspects of classical and celestial mechanics}, in: V. I. Arnold, {\it Dynamical systems III}, (Springer-Verlag, 1999). 

\bibitem{Fom_2004} A. V. Bolsinov and A. T. Fomenko, {\it Integrable Hamiltonian systems}, (Charman and Hall/CRC, 2004). 

\bibitem{Arn_1} V. I. Arnold, \textit{Mathematical methods of classical mechanics},
2nd edn. (Springer, New York, NY, 1989).

\bibitem{AAD23} A. A. Deriglazov, {\it Lagrangian and Hamiltonian formulations of asymmetric rigid body, considered as a constrained system}, 
arXiv:2301.10741. 

\bibitem{AAD23_2} A. A. Deriglazov, {\it Comment on the Letter "Geometric Origin of the Tennis Racket Effect'' by P. Mardesic, et al, Phys. Rev. Lett. 125, 064301 (2020)}, arXiv:2302.04190. 

\bibitem{deriglazov2010classical}
A. A. Deriglazov, {\em Classical mechanics: Hamiltonian and Lagrangian formalism} (Springer, 2nd edition, 2017).

\bibitem{AAD_2022} A. A. Deriglazov, {\it Basic notions of Poisson and symplectic geometry in local coordinates, with applications to Hamiltonian systems}, Universe,  {\bf 8} (2022), 536; 
arXiv:2210.09131. 

\bibitem{AAD_2019}
A. A. Deriglazov, {\it  Nonminimal spin-field interaction of the classical electron and quantization of spin},  Physics of Particles and Nuclei Letters, {\bf 17} 5 (2020) 738-743; arXiv:2001.01294.  

\bibitem{AAD_2021_1} A. A. Deriglazov, {\it Massless polarized particle and Faraday rotation of light in the Schwarzschild spacetime}, 
Phys. Rev. {\bf D 104}, 025006 (2021); arXiv:2103.07794. 

\bibitem{Han_1974} A. J. Hanson and T. Regge, {\it The relativistic spherical top}, Annals of Physics, {\bf 87 (2)} (1974) 498-566. 

\end{thebibliography}
\end{document}